\documentclass[sigconf ,natbib=true,review=false,screen=true]{acmart}

\usepackage{xcolor}
\usepackage{graphicx}
\usepackage{subfigure}
\usepackage{bm}
\usepackage{bbm}
\usepackage{xspace}
\usepackage{url}
\usepackage{multirow}
\usepackage{booktabs}
\usepackage{enumerate}
\usepackage[inline]{enumitem}
\usepackage[skip=0pt]{caption}
\usepackage{acronym}
\usepackage{balance}
\usepackage{mleftright}

\usepackage{algorithm}
\usepackage{algorithmicx}

\usepackage{makecell}

\usepackage{etoolbox}
\usepackage{footmisc}

\newcommand{\wrt}{w.r.t.\xspace}
\newcommand{\ie}{\emph{i.e.,}\xspace}

\newcommand{\eg}{\emph{e.g.,}\xspace}

\newcommand{\ignore}[1]{}

\definecolor{HarrieGreen}{RGB}{34,139,34}

\acrodef{RL}{reinforcement learning}
\acrodef{RL4Rec}{reinforcement learning for recommendation}
\acrodef{Seq-Rec}{sequential recommendation}
\acrodef{RS}{recommender system}
\acrodef{SOFA}{Simulator for OFfline leArning and evaluation}
\acrodef{DANCER}{DebiAsing in the dyNamiC scEnaRio}
\acrodef{EIB}{error-imputation-{\allowbreak}based model}
\acrodef{IPS}{inverse propensity scoring}
\acrodef{DR}{doubly robust}
\acrodef{SNIPS}{self-normalized inverse propensity scoring}
\acrodef{MNAR}{missing not at random}
\acrodef{Pop}{Popularity}
\acrodef{Avg}{average}
\acrodef{MF}{matrix factorization}
\acrodef{BPR}{Bayesian personalized ranking}
\acrodef{TMF}{time-aware matrix factorization}
\acrodef{TTF}{time-aware tensor factorization}
\acrodef{TMTF}{time-aware matrix \& tensor factorization}
\acrodef{MC}{Markov chains}
\acrodef{DQN}{Deep Q-Network}
\acrodef{DQN4Rec}{Deep Q-Network based recommendation}
\acrodef{MDP}{Markov decision process}
\acrodef{GT}{Ground Truth}
\acrodef{sim-GT}{simulated Ground Truth}
\acrodef{CF}{collaborative filtering}
\acrodef{RNN}{recurrent neural network}
\acrodef{GRU}{gated recurrent unit}
\acrodef{CNN}{convolutional neural network}
\acrodef{MLP}{multi-layer perceptron}
\acrodef{LSTM}{long short-term memory}
\acrodef{BOI}{bag of items}
\acrodef{PLD}{pairwise local dependency between items}

\acrodef{NLL}{Negative Log-Likelihood}
\acrodef{PPL}{Perplexity}

\acrodef{P}{Precision}
\acrodef{R}{Recall}
\acrodef{MAP}{Mean Average Precision}
\acrodef{MRR}{Mean Reciprocal Rank}
\acrodef{NDCG}{Normalized Discounted
cumulative gain}

\acrodef{MSE}{Mean Squared Error}
\acrodef{MAE}{Mean Absolute Error}
\acrodef{ACC}{Accuracy}

\makeatletter
\renewcommand\paragraph{\@startsection{paragraph}{4}{\z@}%
                       {-2\p@ \@plus -1\p@ \@minus -1\p@}%
                       {-0.5em \@plus -0.22em \@minus -0.1em}%
                       {\normalfont\normalsize\itshape}}
\makeatother

\author{Jin Huang}
\orcid{0000-0001-9273-9037}
\affiliation{%
\institution{University of Amsterdam}
  \city{Amsterdam}
  \country{The Netherlands}
}
\email{j.huang2@uva.nl}

\author{Harrie Oosterhuis}
\orcid{0000-0002-0458-9233}
\affiliation{%
\institution{Radboud University}
  \city{Nijmegen}
  \country{The Netherlands}
}
\email{harrie.oosterhuis@ru.nl}

\author{Bunyamin Cetinkaya}
\affiliation{%
\institution{University of Amsterdam}
  \city{Amsterdam}
  \country{The Netherlands}
}
\email{bun.cet20@gmail.com}

\author{Thijs Rood}
\affiliation{%
\institution{University of Amsterdam}
  \city{Amsterdam}
  \country{The Netherlands}
}
\email{thijs.rood@hotmail.com}

\author{Maarten de Rijke}
\orcid{0000-0002-1086-0202}
\affiliation{%
\institution{University of Amsterdam}
  \city{Amsterdam}
  \country{The Netherlands}
}
\email{m.derijke@uva.nl}

\renewcommand{\shortauthors}{J. Huang et al.}

\copyrightyear{2022} 
\acmYear{2022} 
\setcopyright{rightsretained} 

\acmConference[SIGIR '22]{Proceedings of the 45th International ACM SIGIR Conference on Research and Development in Information Retrieval}{July 11--15, 2022}{Madrid, Spain.}
\acmBooktitle{Proceedings of the 45th Int'l ACM SIGIR Conference on Research and Development in Information Retrieval (SIGIR '22), July 11--15, 2022, Madrid, Spain}
\acmDOI{10.1145/3477495.3531716}
\acmISBN{978-1-4503-8732-3/22/07}

\usepackage{etoolbox}
\makeatletter
\patchcmd{\maketitle}{\@copyrightpermission}{
  \begin{minipage}{0.3\columnwidth}
    \href{http://creativecommons.org/licenses/by/4.0/}{\includegraphics[width=0.90\textwidth]{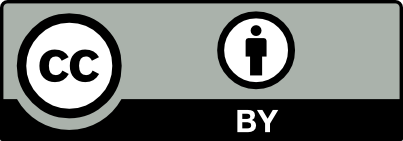}}
  \end{minipage}\hfill
  \begin{minipage}{0.7\columnwidth}
    \href{http://creativecommons.org/licenses/by/4.0/}{This work is licensed under a Creative Commons Attribution International 4.0 License.}
  \end{minipage}

  \vspace{5pt}
}{}{}

\makeatother



\begin{document}


\title{State Encoders in Reinforcement Learning for Recommendation: A Reproducibility Study}

\begin{abstract}
Methods for \ac{RL4Rec} are increasingly receiving attention as they can quickly adapt to user feedback.
A typical \ac{RL4Rec} framework consists of (1)~a state encoder to encode the state that stores the users' historical interactions, and (2)~an \acs{RL} method to take actions and observe rewards.
Prior work compared four state encoders in an environment where user feedback is simulated based on real-world logged user data.
An attention-based state encoder was found to be the optimal choice as it reached the highest performance.
However, this finding is limited to the actor-critic method, four state encoders, and evaluation-simulators that do not debias logged user data.
In response to these shortcomings, we reproduce and expand on the existing comparison of attention-based state encoders (1)~in the publicly available debiased \ac{RL4Rec} \acs{SOFA} simulator with (2) a different \acs{RL} method, (3) more state encoders, and (4) a different dataset.
Importantly, our experimental results indicate that existing findings do \emph{not} generalize to the debiased \acs{SOFA} simulator generated from a different dataset and a \ac{DQN}-based method when compared with more state encoders.
\end{abstract}

\maketitle

\acresetall

\section{Introduction}
\label{sec:intro}
With the development of interactive \acp{RS}, \ac{RL4Rec} is receiving increased attention as  \ac{RL} methods can quickly adapt to user feedback~\citep{afsar-2021-reinforcement,lin-2021-survey}.
\ac{RL4Rec} has been applied in a variety of domains, such as movie~\cite{yuyan2019novel,zhao2019capdrl}, news~\cite{zheng2018drn}, and music recommendations~\cite{pan2019policy}.
A typical process flow of \ac{RL4Rec} starts with an action of the system, which is an item being recommended to the user.
Subsequently, user interactions with the item are returned as feedback (\eg dwell time, a rating, or a click) to the system, which then interprets the feedback as a reward signal.
Finally, with this new interaction, the system updates a state representation that keeps track of the user's historical interactions with the recommended items.
The cycle then repeats as the system again tries to recommend the best item to the user based on its updated state representation.
The goal of \ac{RL4Rec} is to optimize the system so as to achieve the maximum cumulative reward.

An \ac{RL4Rec} framework typically consists of two parts: 
\begin{enumerate*}[label=(\arabic*)]
\item the \emph{state encoder} that encodes the state -- a user's historical interactions -- into a dense representation that is used to estimate the user's preference and the value of state-action pairs; and
\item an \emph{\ac{RL} method} (\eg the \ac{DQN}~\cite{mnih2015human} or the actor-critic~\cite{lillicrap2015continuous} method) that is applied to generate actions based on an estimated state-action value function and observed reward.
\end{enumerate*}
While \ac{RL4Rec} methods have achieved good performance, the effect of the state encoder on \ac{RL4Rec} methods has rarely been explicitly looked at.
To bridge this gap, \citet{liu2020state} compared four state encoders in a simulated \ac{RL4Rec} environment and concluded that an attention-based state encoder leads to the best recommendation performance.
Their findings revealed that the choice of state encoders is important for effective \ac{RL4Rec} and, accordingly, this shows that research into state encoders could further improve the performance of \ac{RL4Rec} methods.
However, the analysis of \citet{liu2020state} is limited to the actor-critic method and only four different state encoders.
Moreover, their evaluation was based on simulated user feedback that was directly inferred from logged user data, which is typically subject to heavy selection bias, \eg popularity bias~\citep{steck2011item}.
Consequently, due to a lack of any bias correction, it is very likely that the results and findings of \citet{liu2020state} are also affected by the selection bias present in the data.

In response to these shortcomings, we reproduce the work by \citet{liu2020state} and generalize its findings concerning state encoders in the following directions:
\begin{enumerate}[leftmargin=*,label=(\arabic*)]
	\item \emph{Different simulated environments}:
	The simulated user feedback used in \citep{liu2020state} is generated from logged user data which is inevitably subject to user selection bias, \eg popularity bias~\citep{pradel2012ranking}.
	Recently, \citet{huang2020keeping} pointed out that simulators that do not debias logged user data yield \ac{RL4Rec} methods that are heavily affected by selection bias.
	Hence, we use SOFA~\citep{huang2020keeping} -- the only publicly available debiased simulator~\cite{bernardi2021simulations} -- to mitigate the effect of selection bias on the resulting \ac{RL4Rec} methods.
	\item \emph{Different RL method}: \ac{DQN} is the most popular \ac{RL} method used in \ac{RL4Rec}~\cite{zhao2020jointly,chen2018stabilizing,zhao2018recommendations,zheng2018drn,sunehag2015deep,liu2018towards,chen2019generative,liu2019deep,yuyan2019novel,qi2019personalized,ie2019reinforcement,liu2020end}; it is structurally simpler than the actor-critic method by only optimizing one objective.
	Thus, it matters to find out whether comparisons of state encoders generalize to \ac{DQN}-based \ac{RL4Rec} methods.
	\item \emph{More state encoders}: Several typical neural networks -- \acp{MLP}, \acp{GRU}, and \acp{CNN} -- %
	are not considered as state encoders by \citet{liu2020state}.
	We expand their comparison by adding these three state encoders based on widely-used typical neural networks.
	\item \emph{Different dataset}: Besides the Yahoo!\ R3 dataset~\cite{marlin2009collaborative} used by \citet{liu2020state}, we also use the Coat shopping dataset~\cite{schnabel2016recommendations} to build the debiased \acused{SOFA}\ac{SOFA} simulator.
\end{enumerate}

\noindent%
We report on our efforts to reproduce the main finding in \citep{liu2020state}:
\begin{quote}
\emph{The attention state encoder for \ac{RL4Rec} provides significantly higher performance than the \ac{BOI}, \ac{PLD} and \ac{Avg} state encoders.}
\end{quote}
Moreover, we investigate whether this finding generalizes in the four directions described above.
Our experimental results show that \citeauthor{liu2020state}'s finding \emph{is} reproducible when applying a \ac{DQN} method and evaluating in the debiased \ac{SOFA} simulator on the Yahoo!\ R3 dataset.
However, we also find that it does \emph{not} generalize to debiased simulations generated from the Coat shopping dataset~\cite{schnabel2016recommendations}.

Our study addresses the following research questions:
\begin{enumerate}[leftmargin=*,label=(RQ\arabic*)] 
	\item Does \citeauthor{liu2020state}'s main finding generalize to the \ac{DQN}-based \ac{RL4Rec} methods when evaluating in the debiased \ac{SOFA} simulator and compared with more state encoders, \ie with the \ac{MLP}, \ac{GRU} and \ac{CNN} state encoders?
	\label{rq:reproduced}
	\item Does \citeauthor{liu2020state}'s main finding generalize to a debiased simulation based on a different dataset?
	\label{rq:dataset}
	\item Should the choice of activation function be taken into account when using the \ac{MLP}-based state encoder for \acp{RL4Rec}?
	\label{rq:activation}
\end{enumerate}

\section{Related Work}

\paragraph{\ac{RL4Rec} methods.}
Deep \ac{RL} methods (\eg \ac{DQN}, actor-critic and REINFORCE) are able to handle high-dimensional spaces and are therefore particularly suitable for \acp{RS} with large state spaces where the user state involves combinatorial user interaction behavior~\cite{afsar-2021-reinforcement}.
\ac{DQN} has been the most popular choice among the \ac{RL4Rec} methods~\cite{zhao2020jointly,chen2018stabilizing,zhao2018recommendations,zheng2018drn,sunehag2015deep,liu2018towards,chen2019generative,liu2019deep,yuyan2019novel,qi2019personalized,ie2019reinforcement,liu2020end}.
\citet{chen2018stabilizing} integrate stratified sampling action replay and approximated regretted rewards with Double \ac{DQN} to stabilize the \ac{RL4Rec} methods in dynamic environments.
\citet{zhao2018recommendations} incorporate positive and negative feedback in a \ac{RL4Rec} method.
\citet{chen2019generative} propose a cascading \ac{DQN} method to obtain a combinatorial recommendation policy with large item space.
\citet{liu2020end} introduce a supervised signal to enable stable training of \ac{RL4Rec} methods.
Others use \ac{DQN} in special recommendation scenarios, \eg for news~\cite{zheng2018drn}, movies~\cite{yuyan2019novel}, education~\cite{liu2018towards}, projects~\cite{qi2019personalized}, slates~\cite{sunehag2015deep,ie2019reinforcement}, or mobile users~\cite{liu2019deep}.
REINFORCE and actor-critic are other two important methods adapted in \ac{RL4Rec}.
REINFORCE is a policy gradient method that directly updates the policy weights~\citep{williams1992simple}. 
\citet{liang2020drprofiling} adapts REINFORCE to find a path between users and items in an external heterogenous information network.
REINFORCE with importance sampling can be used to correct for biases caused by only observing feedback on items recommended by other \acp{RS}~\cite{ma2020off,chen2019top}.
Additionally, REINFORCE is commonly used in conversational \acp{RS}~\cite{greco2017converse,sun2018conversational} and explainable \acp{RS}~\cite{xian2019reinforcement}.
Actor-critic combines REINFORCE and the value-based method~\citep{lillicrap2015continuous}, thus benefiting from both components; it is able to handle large action spaces in \acp{RS}~\cite{dulac2015deep}.
Actor-critic has been used for diverse recommendation tasks~\cite{zhao2018deep,zhao2017deep} and domains~\cite{yu2019vision,zhao2019capdrl}.

In general, \ac{DQN} is the most popular \ac{RL} method used in \acp{RL4Rec} and has a simpler structure than actor-critic. 
Accordingly, we use \ac{DQN} and investigate whether the findings on actor-critic based \ac{RL4Rec} in \citep{liu2020state} generalize to \ac{DQN} based \ac{RL4Rec}.

\paragraph{State encoders.}
Neural networks are widely used in \ac{CF} based recommendation methods; popular choices are \acfp{MLP}~\cite{cheng2016wide,he2017neural}, \acfp{CNN}~\cite{he2018outer}, \acfp{RNN}~\cite{hidasi2015session,wu2017recurrent,yu2016dynamic} and attention~\cite{chen2018sequential,huang2018improving}.
Based on logged user behavior, these methods usually use a neural network to generate a dense vector that captures user preference and can be further used to infer the users' preference over items.
That makes it suitable to adapt these recommendation methods in the state encoders.
Most of the above \ac{RL4Rec} methods use neural networks, such as variants of \acp{RNN}~\cite{chen2019top,zhao2018recommendations}, to construct the state encoder and generate state representation, which can subsequently be used by the previously discussed \ac{RL} methods.
However, the effect of state encoders has rarely been explored explicitly.
To the best of our knowledge, \citet{liu2020state} are the first to compare the effects of different state encoders in \ac{RL4Rec} methods.
We continue this research direction by reproducing and generalizing \citeauthor{liu2020state}'s comparison.

\paragraph{Debiasing recommendations.}
Bias is prevalent in interactions with \acp{RS}, such as users choosing to rate certain items more often (self selection bias)~\cite{pradel2012ranking, steck2010training} and \acp{RS} showing certain items to users more often (algorithmic selection bias)~\cite{hajian2016algorithmic}.
As a result, user preference prediction may be biased and over-specialization~\cite{adamopoulos2014over}, consequently, filter bubbles~\cite{pariser2011filter,nguyen2014exploring} and unfairness~\cite{chen2020bias} may occur.
To correct for bias, debiasing methods may be applied, such as the error-imputation-based method~\cite{steck2010training}, \ac{IPS}~\citep{horvitz1952generalization}, and the doubly robust method~\citep{robins1994estimation, kang2007demystifying}.
\ac{IPS} is the most popular method and widely used in debiasing recommendations~\cite{chen2019correcting,joachims2017unbiased,schnabel2016recommendations,chen2019top,ma2020off}.
Corrections of the debiased methods may lead to substantially improved prediction performance~\cite{schnabel2016recommendations}.

\paragraph{RL4Rec simulators.}
The usage of simulated \ac{RL4Rec} environments is widespread~\cite{bernardi2021simulations,zheng2018drn,li2010contextual,rohde2018recogym,shi2019virtual,shi2019pyrecgym,ie2019recsim,zhao2019toward,zhang2020evaluating} and for a good reason:
\ac{RL4Rec} methods learn by directly interacting with users but the online nature of this learning process brings risks and limitations:
\begin{enumerate*}[label=(\arabic*)]
\item in practice, the user experience can be negatively affected during the early stages of the learning process; and 
\item research and experimentation with \ac{RL4Rec} systems is often infeasible since most researchers have no access to real interactions with live users.
\end{enumerate*}
\Ac{RS} simulators mitigate these issues as they allow \ac{RS} developers and researchers to optimize and evaluate their \ac{RL4Rec} methods on simulated user behavior~\cite{bernardi2021simulations,rohde2018recogym,shi2019pyrecgym,ie2019recsim}.
Some simulators generate user behavior based on fully synthetic data (\eg generated from a Bernoulli distribution~\cite{rohde2018recogym}).
These have been critiqued for oversimplifying user behavior~\cite{shi2019pyrecgym,bernardi2021simulations}.
Alternatively, to match real user behavior more closely, other simulators generate user behavior based on logged user data~\cite{shi2019pyrecgym,ie2019recsim,zhao2019toward}.
While these simulators are widely accessible, most ignore the interaction biases present in the logged user data from which they generate simulated user behavior.
Recently, \citet{huang2020keeping} have pointed out that simulators that do not debias logged user data result in \ac{RL4Rec} models that are also heavily affected by the selection biases.
They argue that, as a result, findings based on the outcomes of such biased simulators can be misleading because the effect of the interaction biases extend to the results underlying such findings.
To mitigate the effect of bias, the \ac{SOFA} environment~\cite{huang2020keeping} applies \acf{IPS} to reduce selection bias in logged user data when learning user preference and thus provides a debiased simulator.
To the best of our knowledge, \ac{SOFA} is the only publicly available debiased simulator.
Therefore, we use \ac{SOFA} to train and evaluate \ac{RL4Rec} methods with different state encoders.

\section{Preliminaries -- RL4Rec}
\label{sec:preliminaries}

\ac{RL4Rec} methods commonly model the recommendation task as a \acf{MDP}, where optimization is based on interactions between the \ac{RS} (\ie the agent) and users (\ie the environment).
The elements of an \ac{MDP} for \ac{RL4Rec} are:
\begin{description}[leftmargin=\parindent,nosep]
    \item[State space] $\mathcal{S}$: A state $s^u_t$ stores the interaction history of user $u$ at $t$-th turn. For clarity and brevity, we omit the superscript $u$ when the user is clear from the context. The state $s_t$ consists of the items recommended by the \ac{RS} and the corresponding user feedback (\eg click or skip), denoted as $s_t = ([i_1, i_2, \ldots, i_t]$, $[f_1$, $f_2$, \ldots, $f_t])$. In turn $t+1$, the \ac{RS} takes an action based on the information represented in state $s_t$. The state $s^u_0$ is always initialized as empty, denoted as $s^u_0=([\ ], [\ ])$.
    \item[Action space] $\mathcal{A}$: The action $a_t$ is to recommend an item $i_t$ to user $u$ by the \ac{RS} based on state $s_{t-1}$ in turn $t$. Similar to the setup of \citet{liu2020state}, in the \ac{SOFA} simulator the \ac{RS} only recommends one item to the user at every turn.
    \item[Reward] $\mathcal{R}$: The immediate reward $r(s_{t-1}, a_t)$ is generated according to user's feedback $f_t$ (\eg skip or click) on $a_t$.
    \item[Transition probability] $\mathcal{P}$: In turn $t+1$, \ac{SOFA} receives an item $i_{t+1}$ being recommended from the \ac{RS} and assumes that the state $s_{t}$ transitions deterministically to the next state $s_{t+1}$ by appending item $i_{t+1}$ and the corresponding user feedback $f_{t+1}$, denoted as $s_{t+1} = ([i_1, i_2, \ldots, i_{t+1}], [f_1, f_2, \ldots, f_{t+1}])$.
    \item[Discount factor] $\gamma$: $\gamma \in [0, 1]$ determines the degree to which the \ac{RS} cares about future rewards: if $\gamma=0$, the \ac{RS} only takes the immediate reward into account when taking an action; if $\gamma=1$, the sum of all future rewards is considered.
\end{description}

\begin{figure}[t]
    \centering
    \includegraphics[width=1.0\linewidth]{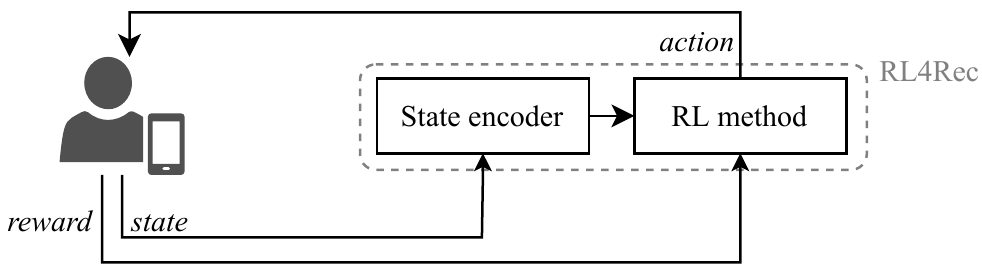}%
    \caption{The general framework of \ac{RL4Rec}.}
    \label{fig:RL4Rec}
\end{figure}

\label{sec:RL4Rec}
\noindent%
Generally, the \ac{RL4Rec} method includes two components as shown in Fig.~\ref{fig:RL4Rec}:
\begin{enumerate*}[label=(\arabic*)] 
    \item the state encoder is applied to encode a state $s$ into a dense representation that captures the user preference and is subsequently used to approximate the state-action value function $\widehat{Q}(s, a\,;\, \theta)$; for every action $a\in \mathcal{A}$, $\widehat{Q}(s, a; \theta)$ represents the expected reward following the recommendation of item $a$ in state $s$; and
    \item the \ac{RL} method decides which action to take based on the state representation, and chooses how the parameters of the policy and state encoder models should be updated according to the rewards received from the user.
\end{enumerate*}

While the \ac{RL} method chooses items to recommend to the user, it bases its decisions on the state representations provided by the state encoder.
Therefore, the performance of an \ac{RL4Rec} system heavily relies on the functioning of the state encoder.
As a result, understanding how the choice of state encoder should be made is central to \ac{RL4Rec}.

\section{Original State Encoder Comparison}
\label{sec:original-setting}

\citet{liu2020state} follow the \ac{RL4Rec} framework detailed in Section~\ref{sec:RL4Rec} and apply an actor-critic \ac{RL} method to take actions and update the model parameters.
The applied actor-critic method comprises two components:
\begin{enumerate*}[label=(\arabic*)]
    \item the actor network follows the policy $\pi_{\theta^A}(s_{t-1}) \in \mathbb{R}^{d}$ and takes the action $a_t$, \ie to recommend item $i$ with the maximum ranking score $\bm{q}_i^\top \pi_{\theta^A}(s_{t-1})$, where $\bm{q}_i$ denotes the embedding of item $i$;
    \item the critic network estimates state-action value function $\widehat{Q}(s, a; \theta^C)$ as the approximation of the true state-action value function that represents the merits of the recommendation policy generated by the actor network.
\end{enumerate*}
The target network technique is also adopted, where an identical actor network with policy $\pi_{\theta^{A'}}$ and an identical critic network with state-action value function $\widehat{Q}(s,a;\ \theta^{C'})$ are used. 
The recommendation agent makes use of experience replay and employs a replay memory $\mathcal{D}$ to store the agent's experience, \ie the user interactions with the recommended items in the \ac{RL4Rec} domain.
Given transitions $(s_{t-1}, a_t, r_t, s_t) \in \mathcal{D}$ generated based on the interactions between the user and recommendation policy $\pi_{\theta^A}$,
\ie $a_t \sim \pi_{\theta^A}(s_{t-1})$,
the parameters $\theta^A$ of the actor network and $\theta^C$ of the critic network are updated as:
\begin{equation}
\begin{split}
    \theta^A & \gets \theta^A + \alpha^A \widehat{Q}(s_{t-1}, a_t; \theta^C)\nabla_{\theta^A}\log \pi_{\theta^A}(s_{t-1}), \\
    \theta^C & \gets \theta^C + \big(\alpha^C(r_t+\gamma \widehat{Q}(s_t, a_{t+1}; \theta^{C'})
    \\&\hspace{2.15cm}
    -\widehat{Q}(s_{t-1}, a_t; \theta^C))\nabla_{\theta^C}\widehat{Q}(s_{t-1}, a_t; \theta^C)
    \big),
   \end{split}   
\end{equation}
where $\alpha^A$ and $\alpha^C$ denote the learning rates for the actor network and the critic network, respectively, and $a_{t+1} \sim \pi_{\theta^{A'}}(s_t)$.
The target network is updated following the soft replace technique: given a soft-replace parameter $\tau$, the parameters $\theta^{A'}$ of the actor network and $\theta^{C'}$ of the critic network are updated as follows:
\begin{equation}
    \theta^{A'} \gets \tau \theta^A + (1 - \tau)\theta^{A'}, \qquad
    \theta^{C'} \gets \tau \theta^C + (1 - \tau)\theta^{C'}.
\end{equation}
\citet{liu2020state} consider two types of state encoder methods for representing states and approximating the state-action value functions, with and without user embedding $\bm{p}_u$.
First, they introduce an item-to-item \acl{CF} method, DRR-p, without taking user embeddings into account, which uses an element-wise product to capture the pairwise local dependency between items:
\begin{align}
    \bm{s}_t = [\bm{q}_{i_1}, \bm{q}_{i_2}, \ldots, \bm{q}_{i_t}, \{w_i \bm{q}_i \otimes w_j \bm{q}_j \mid i, j \in \{i_1, i_2, \ldots, i_t\} \}],
\end{align}
where $\otimes$ denotes the element-wise product; and the scalars $w_i$ and $w_j$ indicate the importance weights of items $i$ and $j$, respectively.
Additionally, three state encoders, DRR-u, DRR-ave and DRR-att, with user embeddings are introduced and outperform the state encoder DRR-p without user embeddings:
\begin{enumerate*}[label=(\arabic*)]
    \item the element-wise product on user-item embedding pairs is incorporated:
    $
    \bm{s}_t = [\{\bm{p}_u \otimes w_i \bm{q}_i \mid{}$ $i \in\{i_1, i_2, \ldots, i_t\}\}];
    $
    and
    \item to reduce computational costs, the weighted average pooling schema is used to aggregate the item embeddings: $\bm{s}_t = [\bm{p}_u, \bm{p}_u \otimes \{ \text{ave}(w_i \bm{q}_i) \mid i \in\{i_1, i_2, \ldots, \\ i_t\}\}]$, where $\text{ave}(\cdot)$ denotes the average pooling operator. %
   Finally,  \item an attention network is applied:
\end{enumerate*}
\begin{align}
\bm{s}_t & = [\bm{p}_u, \bm{p}_u \otimes \{ \text{ave}(a_{u,i} \bm{q}_i) \mid i \in\{i_1, i_2, \ldots, i_t\}\}],\\
    a_{u,i} & = \frac{\exp(a_{u,i}')}{\sum_{i'\in \{i_1, i_2, \ldots, i_t \}}\exp(a_{u,i'}')}, \\
     a_{u,i}' & = \text{ReLU}(([\bm{p}_u, \bm{q}_i] \bm{W}_2)+b_2)\bm{W}_1 + b_1,
\end{align}
where the weight matrices $\bm{W}_1, \bm{W}_2$ and the bias vectors $b_1, b_2$ project the input into a hidden layer; ReLU is the activation function for the hidden layer.

Given the state representation $\bm{s}_t$, the ranking score of item $i$, \ie $\bm{p}_i^\top\pi_{\theta^A}(s_{t})$, can be used to execute policy $\pi_{\theta^A}(s_{t})$ and approximate state-action value function $\widehat{Q}(s_t, a; \theta^C)$. 
Consequently, the resulting actor-critic based \ac{RL4Rec} method can interact with the (simulated) users and update the parameters iteratively.
\citet{liu2020state} compare the actor-critic based \ac{RL4Rec} method with four state encoders, DRR-p, DRR-u, DRR-ave and DRR-att, in simulators generated from two datasets~\cite{harper2015movielens} containing temporal information and two datasets not containing temporal information~\cite{goldberg2001eigentaste,marlin2009collaborative}.
They conclude that: 
\begin{enumerate*}[label=(\arabic*)]
    \item state encoders that utilize user embeddings outperform state encoders without user embeddings;
    \item the average pooling schema can decrease the dimensionality of the state representation to reduce overfitting and improve recommendation performance; and
    \item the attention-based state encoder provides the best performance among the four state encoders introduced above.
\end{enumerate*}

\section{Our Reproduced State Encoder Comparison}
\label{sec:reproduce:comparison}

Having specified the setting of \citet{liu2020state}'s study (cf.\ Section~\ref{sec:original-setting}), we generalize \citeauthor{liu2020state}'s finding to four directions and can summarize the following key differences:
\begin{enumerate*}[label=(\arabic*)]
    \item \textbf{Different simulated environments}: We adopt \ac{SOFA}, the debiased simulator, which mitigates the effect of bias present in logged data when generating user preferences on items; in contrast, \citet{liu2020state} use a simulation directly generated from logged data without considering bias.  
    \item \textbf{Different \ac{RL} method}: We apply \ac{DQN}, which is widely used in \ac{RL4Rec} research and has a simpler structure with optimizing only one objective, whereas \citet{liu2020state} apply the actor-critic method.
    \item \textbf{More state encoders}: Besides the four state encoders proposed by~\citet{liu2020state}, we expand the comparison by adding three more state encoders based on typical neural network architectures: \ac{MLP}, \ac{GRU} and \ac{CNN}, 
    which are widely used in recommendation methods to generate representations according to historical user interactions.
    \item \textbf{Different dataset}:
    Our comparison uses the \text{Yahoo!\ R3}, which is also used by~\citet{liu2020state}, as well as the Coat shopping dataset~\cite{schnabel2016recommendations}, which is not considered in \citep{liu2020state}.
    To the best of our knowledge, these two datasets are the only publicly available datasets that can be used to unbiasedly simulate  recommendations, since part of their data is gathered on randomized recommendations.
    Unfortunately, there are two more datasets used by~\citet{liu2020state} that cannot be used in \ac{SOFA} due to a lack of randomized data for debiasing.
\end{enumerate*}

It is crucial to understand whether the choice of state encoder is important, and if so, what factors should be considered when making this choice.
In particular, the aim of our reproducibility study is to analyze whether the choice of state encoder is robust \wrt the effect of bias, the choice of \ac{RL} method, and the sources of data used.
The differences listed above allow us to address this aim and investigate whether the findings of \citet{liu2020state} generalize along these dimensions.

Below, we describe the setting in which we reproduce and expand on the comparisons performed by \citet{liu2020state}.
Section~\ref{sec:sofa} details the debiased \ac{SOFA} simulator that we use, 
Section~\ref{sec:DQN4Rec} explains the \ac{DQN} \ac{RL} method that is applied,
and finally, Section~\ref{sec:reproduce:stateencoders} lists the state encoders included in our comparison.

\subsection{\acf{SOFA}}
\label{sec:sofa}

To mitigate the effect of bias present in logged data, \citet{huang2020keeping} propose a debiased simulator, named \ac{SOFA}, which consists of two components:
\begin{enumerate*}[label=(\arabic*)]
    \item a debiased user-item rating matrix to present user preferences for items, and
    \item a user choice model to simulate user feedback and generate the next state and the immediate reward.
\end{enumerate*}
The bias mitigation step is applied between the logged data and the learned user preference prediction model, thereby mitigating the bias originating from the logged data from affecting the user preference prediction model.
User behavior (\eg ratings) could be affected by various forms of selection bias, 
\eg users tend to rate more popular items (\emph{popularity bias})~\cite{pradel2012ranking,steck2011item} or the items that they expect to enjoy beforehand (\emph{positivity bias})~\cite{pradel2012ranking}.
This is generally modelled by decomposing the probability of observing a rating $y_{u,i}$ given by user $u$ on item $i$ into (1) the preference $P(y_{u,i})$, \ie the distribution over rating values the user $u$ would give to item $i$; and (2)
the propensity $P(o_{u,i})$, \ie the probability of observing any rating from user $u$ for item $i$ in the dataset.
The assumed model is thus: 
\begin{equation}
    P(o_{u,i}, y_{u,i}) = P(o_{u,i})P(y_{u,i}),
\end{equation} 
where $o_{u,i}$ denotes the observation indicator: $o_{u,i}=1$ if the rating $y_{u,i}$ is observed,  otherwise, $o_{u,i}=0$ indicates a rating is missing.
Due to bias, certain ratings are more likely to be observed than others.
In other words, $P(o_{u,i})$ is not uniform over all user-item pairs.
As a result, naively ignoring the propensities during evaluation or optimization gives more weight to the user-item pairs that are overrepresented due to bias~\citep{schnabel2016recommendations}, \eg giving the most weight to the most popular items.
In turn, this results in biased user rating predictions $\hat{y}_{u,i}$ that fail to match the true user ratings $y_{u,i}$.
The bias mitigation step of \ac{SOFA} applies \acf{IPS}~\cite{imbens2015causal} to inversely weight ratings according to the corresponding observation probabilities so that, in expectation, each user-item pair is represented equally.
Let $\delta(y_{u,i}, \hat{y}_{u,i})$ indicate the loss resulting from the match between true rating $y_{u,i}$ and the predicted rating $\hat{y}_{u,i}$~\cite{schnabel2016recommendations}:
\begin{equation}
\mbox{}\!
    \mathbb{E}\!\left[\mathcal{L}_\text{IPS}\right] \propto \mathbb{E}\!\left[\frac{\delta(y_{u,i}, \hat{y}_{u,i})}{P(o_{u,i}=1)}\!\right] 
    \!=\! \frac{\mathbb{E}[o_{u,i}] \delta(y_{u,i}, \hat{y}_{u,i})}{P(o_{u,i}=1)}
    \!=\! \delta(y_{u,i}, \hat{y}_{u,i}).
    \label{eq:IPS}
\end{equation}
Therefore, using the \ac{IPS} debiasing method, \ac{SOFA} can learn debiased user preferences for items and mitigate the effect of bias on the resulting simulated user behavior and the final produced \ac{RL4Rec} methods~\citep{huang2020keeping}.

Before the interaction starts, \ac{SOFA} uniformly randomly samples a batch of users and initializes their states as empty;
then \ac{SOFA} interacts with \ac{RL4Rec} methods over ten turns.
Within \ac{SOFA}, the \ac{RL4Rec} methods aim to maximize the cumulative number of clicks received over ten interaction turns, and are accordingly evaluated on the cumulative number of clicks they receive over ten interaction turns.
Furthermore, \ac{SOFA} provides a general \ac{DQN}-based \ac{RL4Rec} framework, which Section~\ref{sec:DQN4Rec} describes in detail.

\subsection{Deep Q-Network based recommendation}
\label{sec:DQN4Rec}

\acfp{DQN}~\cite{mnih2015human} are based on Q-learning, one typical value-based \ac{RL} method~\citep{sutton1998reinforcement}, while the actor-critic methods integrate a value-based method with the policy gradient REINFORCE method~\citep{williams1992simple}.
As a result, \acp{DQN} have a simpler structure than actor-critic methods by only optimizing one objective;
thus, while actor-critic methods are potentially more powerful for handling large state and action spaces, \acp{DQN} can be more data-efficient.
\acp{DQN} have been widely used in \ac{RL4Rec} to improve recommendation performance~\cite{zhao2020jointly,chen2018stabilizing,zhao2018recommendations,zheng2018drn,sunehag2015deep,liu2018towards,chen2019generative,liu2019deep,yuyan2019novel,qi2019personalized,ie2019reinforcement,liu2020end}.
For these reasons, we follow \ac{SOFA} and choose to use the basic \ac{DQN} for our reproducibility study.
We optimize the \ac{DQN} by fitting its predicted state-action function $\widehat{Q}(s, a; \theta)$ to the expected discounted cumulative reward $\sum_t\gamma^t r_t$.
To stabilize the training process, \ac{DQN} introduces a behavior network separate from the target network. 
Here, we apply a state encoder as the behavior network and an identical state encoder as the target network.
These two state encoders have the same structure and use the same item embeddings, but are updated in different ways.
Moreover, \ac{DQN} makes use of experience replay and employs a replay memory $\mathcal{D}$ to store the agent's experience, \ie the user interactions with the recommended items in the \ac{RL4Rec} domain.

Given a transition $(s_{t-1}, a_t, r_t, s_{t}) \in \mathcal{D}$,
the behavior network estimates Q-value function $\widehat{Q}(s_{t-1}, a_{t}; \theta)$ on the given state-action pair $(s_{t-1}, a_t)$, where $\theta$ denotes the parameters of the behavior network; the target network is used to estimate Q-value function $\widehat{Q}'(s_t, a; \theta')$ for any action $a\in \mathcal{A}$ given state $s_t$, with the parameters $\theta'$ fixed and periodically copied from $\theta$ in the behavior network.
Following~\cite{mnih2015human}, the parameters $\theta$ of the behavior network are updated by minimizing the following smooth L1 loss function for steady gradients with the Adam optimizer:
\begin{equation}
        \mathcal{L}(\theta) = \mathbb{E}_{(s_{t-1}, a_t, r_t, s_t)\sim D}
        \begin{cases}
            0.5 (\delta^\text{TD})^2 & \text{if $|\delta^\text{TD}| < 1$,}\\
            |\delta^\text{TD}| & \text{otherwise.}
        \end{cases} 
\end{equation}
\begin{equation}
        \delta^{\text{TD}} = r_t+\gamma \max_{a}\widehat{Q}'(s_t, a; \theta') - \widehat{Q}(s_{t-1}, a_t; \theta).
\end{equation}
Note that the parameters $\theta'$ of the target network are not updated in each learning step, but periodically replaced by $\theta$ after multiple learning steps.

\subsection{State encoders in our comparison}
\label{sec:reproduce:stateencoders}

As described in Section~\ref{sec:preliminaries}, the state encoder is used to generate representations of the state that can be used as input to the approximated state-action value function.
The choice of state encoder can have a large impact on the performance of the \ac{RL4Rec} system~\citep{liu2020state}.
Accordingly, it is crucial to select an appropriate and effective state encoder.
Since \citet{liu2020state} have not made their source code publicly available, we have reimplemented the four state encoders of their original comparison (see Section~\ref{sec:original-setting}): DRR-p, DRR-u, DRR-ave and DRR-att.
Due to the increasing importance of privacy and the fact that \ac{SOFA} does not provide user information, we drop the user embedding \wrt the user id and add user feedback to the recommended items to obtain user preferences in these four state encoders, renamed as \acf{PLD}, \acf{BOI}, \acf{Avg} and Attention.
Additionally, we consider three more typical neural networks -- \ac{MLP}, \ac{GRU} and \ac{CNN} -- when constructing the state encoders.

We use $\bm{q}_i$ to denote the embedding of item $i$ and $\bm{f}_i$ for the embedding of feedback $f_i \in \{0, 1\}$ from the user on the item $i$.
Given state $s_t = ([i_1, i_2, \ldots, i_t], [f_1, f_2, \ldots, f_t])$, we have the corresponding item embeddings $[\bm{q}_{i_1}, \bm{q}_{i_2}, \ldots, \bm{q}_{i_t}]$ and feedback embeddings $[\bm{f}_{i_1}, \bm{f}_{i_2}, \ldots, \bm{f}_{i_t}]$.
The state-action value function $\widehat{Q}(s_t, a)$ can be approximated by the following state encoders:
\begin{description}[leftmargin=\parindent]%
    \item[\Ac{BOI}:] Corresponding to DRR-u from~\citet{liu2020state}, 
    the state representation $\bm{s}^\text{BOI}_t$ is formulated as a list of weighted element-wise products of historical item embeddings and the corresponding feedback embeddings. Then, one linear layer is applied and the dimensionality of the output space is set to the number of items:
    \begin{equation}
    \begin{split}
        \bm{s}^\text{BOI}_t &= [\{w_{i}\bm{q}_{i} \otimes \bm{f}_{i} \mid i\in\{i_1, i_2, \ldots i_t\}\}], \\
        \widehat{Q}(s_t, a) &= \bm{W}^\top \bm{s}^\text{BOI}_t + b.
    \end{split}
    \end{equation}
    \item[\Ac{PLD}:] Corresponding to DRR-p from~\citet{liu2020state}, the \acl{PLD} $e_{i,j}$ is also considered in modeling state representation $\bm{s}^\text{PLD}_t$:
    \begin{equation}
    \begin{split}
        \bm{s}^\text{PLD}_t &= [\{w_{i}\bm{q}_{i} \otimes \bm{f}_{i} \mid i\in\{i_1, i_2, \ldots i_t\}\},
        \\&\hspace{1.28cm}
         \{e_{i,j}\mid i, j \in \{i_1, i_2, \ldots, i_t\} \}], \\
        e_{i,j} &= w_i (\bm{q}_i\otimes \bm{f}_i)^\top (\bm{q}_j\otimes \bm{f}_j)w_j, \\
        \widehat{Q}(s_t, a) &= \bm{W}^\top \bm{s}^\text{PLD}_t + b.
    \end{split}
    \end{equation}
    \item[\ac{Avg}:] Corresponding to DRR-ave from~\citet{liu2020state}, one linear layer is applied with no activation function and the dimensionality of the output space is set to the number of items:
    \begin{equation}
        \widehat{Q}(s_t, a) = \bm{W}^\top \text{ave}(\{\bm{q}_i\otimes \bm{f}_i | i \in \{i_1, i_2, \ldots, i_t\}\}) + b,
    \end{equation} 
    where $\text{ave}(\cdot)$ denotes the component-wise average operator on a set of vectors; and $\bm{W}$ and $b$ are the weight and bias term of the linear layer, respectively.
    \item[\ac{MLP}:] Novel in our comparison, on top of \ac{Avg}, we lift the linear assumption of state and state-action value function by applying a non-linear activation function $\sigma$, \eg tanh, ReLU, or sigmoid:
    \begin{equation}
        \widehat{Q}(s_t, a) = \sigma(\bm{W}^\top \text{ave}(\{\bm{q}_i\otimes \bm{f}_i | i \in \{i_1, i_2, \ldots, i_t\}\}) + b).
    \end{equation}
    \item[\ac{CNN}:] Novel in our comparison, a basic \ac{CNN} with one convolution layer and one max-pooling layer is applied; to compute $\widehat{Q}(s_t, a)$, a fully-connected layer is also adopted with the dimensionality being the number of items:
    \begin{equation}
        \widehat{Q}(s_t, a) = \bm{W}^\top \text{max}(W_{C}(([\bm{q}_{i_1}, \ldots, \bm{q}_{i_t}, \bm{f}_{i_1}, \ldots, \bm{f}_{i_t}]^\top))) + b,\!\!\!\mbox{}
    \end{equation}
    where $\text{max}(\cdot)$ denotes the max operator of the max-pooling layer; $W_C$ indicates the weight function of a l-dilated convolution filter of size $3\times 3$ and the activation function ReLU; and $\bm{W}$ and $b$ are the weight and bias term of the fully-connected layer, respectively.
    \item[\ac{GRU}:] Novel in our comparison, a basic \ac{GRU} layer and a dense layer are applied:
    \begin{equation}
    \begin{split}
        \bm{h}_k &= W_G(\bm{h}_{k-1}, \bm{q}_{i_k} \otimes \bm{f}_{i_k}), \quad \forall k=1, 2, \ldots, t  \\
        \widehat{Q}(s_t, a) &= \bm{W}^\top \bm{h}_t + b,
    \end{split}
    \end{equation}
    where $W_G$ indicates the weight function of the \ac{GRU} unit with the activation funtion tanh; and $\bm{h}_0$ is set as a zero-vector. The hidden state vector $\bm{h}_k$ is computed conditioned on the previous hidden state vector $\bm{h}_{k-1}$ and the input $\bm{q}_{i_k} \otimes \bm{f}_{i_k}$.
    \item[Attention:] Corresponding to DRR-att from~\citet{liu2020state}, following~\cite{bahdanau2014neural} we insert an attention layer into the \ac{GRU}-based state encoder: 
    \begin{align}
        a_k &= \frac{\exp(a'_k)}{\sum^t_{k'=1}\exp(a'_{k'})}, \quad a'_k = (\bm{W}_A^\top \bm{h}_t)^\top \bm{h}_k,\\
        \label{eq:att}
        \widehat{Q}(s_t, a) &= \bm{W}^\top\left[\left(\sum^t_{k=1}a_k \bm{h}_k\right), h_t\right] + b,
    \end{align}
    where $W_A$ denotes the weight function of the attention layer; $a_k$ denotes the attention weight on the hidden state vector $\bm{h_t}$; and the attentive combination of all the hidden state vectors is used to compute the state-action value function $\widehat{Q}(s_t, a)$.
\end{description}

\section{Experimental Setup}
\label{sec:experiment-setup}

In this section, we describe the experiments performed to answer the research questions presented in Section~\ref{sec:intro}.

\paragraph{Datasets and simulators.} We use \ac{SOFA} to generate two debiased simulations that simulate user behavior based on two real-world datasets: Yahoo!\ R3~\cite{marlin2009collaborative} and Coat shopping~\cite{schnabel2016recommendations},
which -- to the best of our knowledge -- are the only publicly available datasets that include a uniformly-random sampled test-set that allows for unbiased evaluation.
The number of users in the Yahoo!\ R3 and Coat shopping datasets are 15,400 and 290, respectively; and the number of items are 1,000 and 300, respectively.
Both datasets include a biased training set and an unbiased test set:
the training set contains ratings observed from \emph{natural} real-world user behavior, whereas the test set contains ratings asked from users on uniformly randomly sampled items.
Consequently, the training set is affected by the forms of bias present in standard user interactions, but the test set is unaffected by any selection bias since it relies on uniform random sampling.
The simulations used for training \ac{RL4Rec} methods are based on debiased user preferences generated from \ac{IPS}-based rating prediction methods (Eq.~\ref{eq:IPS}) on the biased training set; in contrast, the evaluation of the \ac{RL4Rec} methods is performed on the unbiased simulations generated from the unbiased test sets.

\paragraph{Hyperparameters.}
The required hyperparameters come in two kinds:
\begin{enumerate*}[label=(\arabic*)] 
    \item Hyperparameters of the used \ac{DQN}: we follow the hyperparameters reported by~\citet{huang2020keeping} (see Table~\ref{tab:DQN:hyperparameters}) and fix the values for the \ac{DQN} based \ac{RL4Rec} methods with different state encoders.
    \item Hyperparameters used in the state encoders:  the common hyperparameters are tuned per state encoder in the following ranges: learning rate $\eta \in \{10^{-5}, 10^{-4}, 10^{-3}\}$ and the dimension of item embedding $d \in \{16,\ 32,\ 64\}$. Additionally, the dimensions of the weight functions in the \ac{CNN}, \ac{GRU} and attention state encoders are taken from $d' \in \{16,\ 32,\ 64\}$.
\end{enumerate*}

\begin{table*}[t]
	\centering
        \caption{List of hyperparameters for \ac{DQN} and their values.} \vspace{1mm}
		\label{tab:DQN:hyperparameters}%
			\begin{tabular}{p{0.55\columnwidth} p{1.3\columnwidth} r}
			\toprule
            \bf Hyperparameter & \bf Definition & \bf Value \\
			\midrule
			Memory Size & The number of transitions stored in the replay memory. & 6,000 \\
            Discount factor & Discount factor $\gamma$ used in the \ac{DQN}. & 0.9 \\
            Epsilon & The minimal probability of recommending an item randomly when taking an action.  & 0.1\\
            Epsilon decay frequency & The number of step with which the epsilon $\epsilon$ (initial value as 0.8) minus 0.1. & 20,000\\
            Minibatch size & The number of training cases randomly selected from replay memory and being used to update the parameters of policy. & 128 \\
            Targetnet replacement frequency & The number of step with which the target network is updated. & 20\\
			\bottomrule
		\end{tabular}%
\end{table*}%

\paragraph{Evaluation metrics.} 
As introduced in Section~\ref{sec:sofa} we use the cumulative number of clicks received over 10 interaction turns in the unbiased simulated online environments to evaluate the performance of the state encoders in the \ac{DQN4Rec} method.
The cumulative or average number of clicks is a common choice of metric~\cite{zhao2018recommendations,liu2020state} for online evaluation of \ac{RL4Rec} since it can indicate the long-term user engagement performance achieved by \acp{RL4Rec}.

\paragraph{Release of implementation.}
The complete implementation of our experiments with accompanying documentation and additional resources are publicly available for future reproducibility at \url{https://github.com/BetsyHJ/RL4Rec}.

\begin{figure*}[!ht]
    \centering
	\subfigure{
		\includegraphics[width=0.7\textwidth,trim=0 0 0 0]{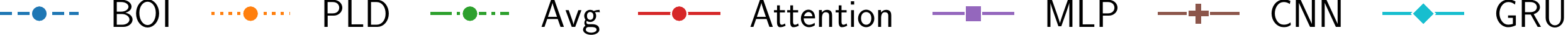}
	} \setcounter{subfigure}{0}
	\subfigure[Yahoo!\ R3.]{
	  \includegraphics[width=.47\textwidth]{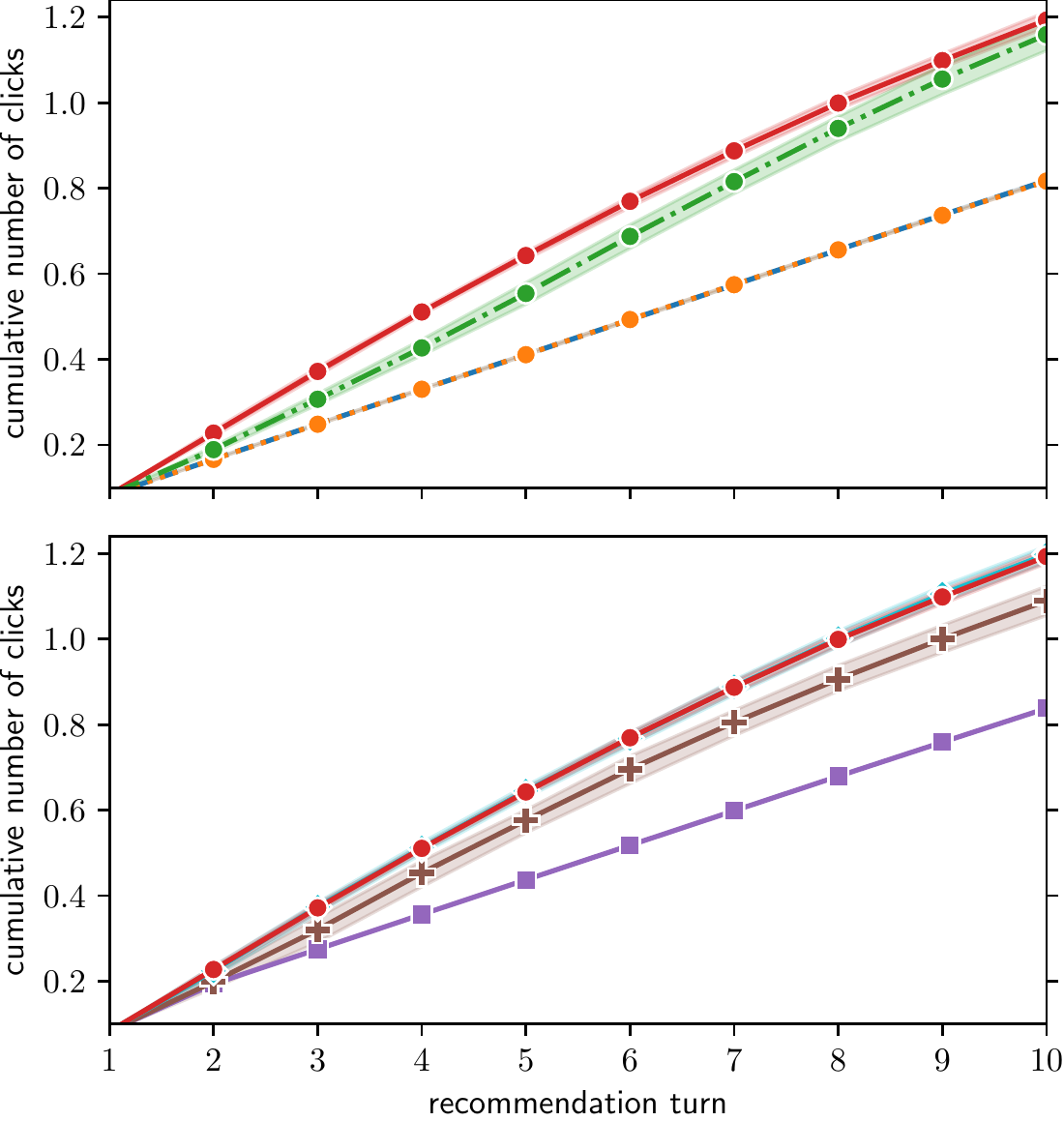}
	  \label{fig:eval_curves:yahoo}
	  } %
    \subfigure[Coat shopping.]{
	  \includegraphics[width=.47\textwidth]{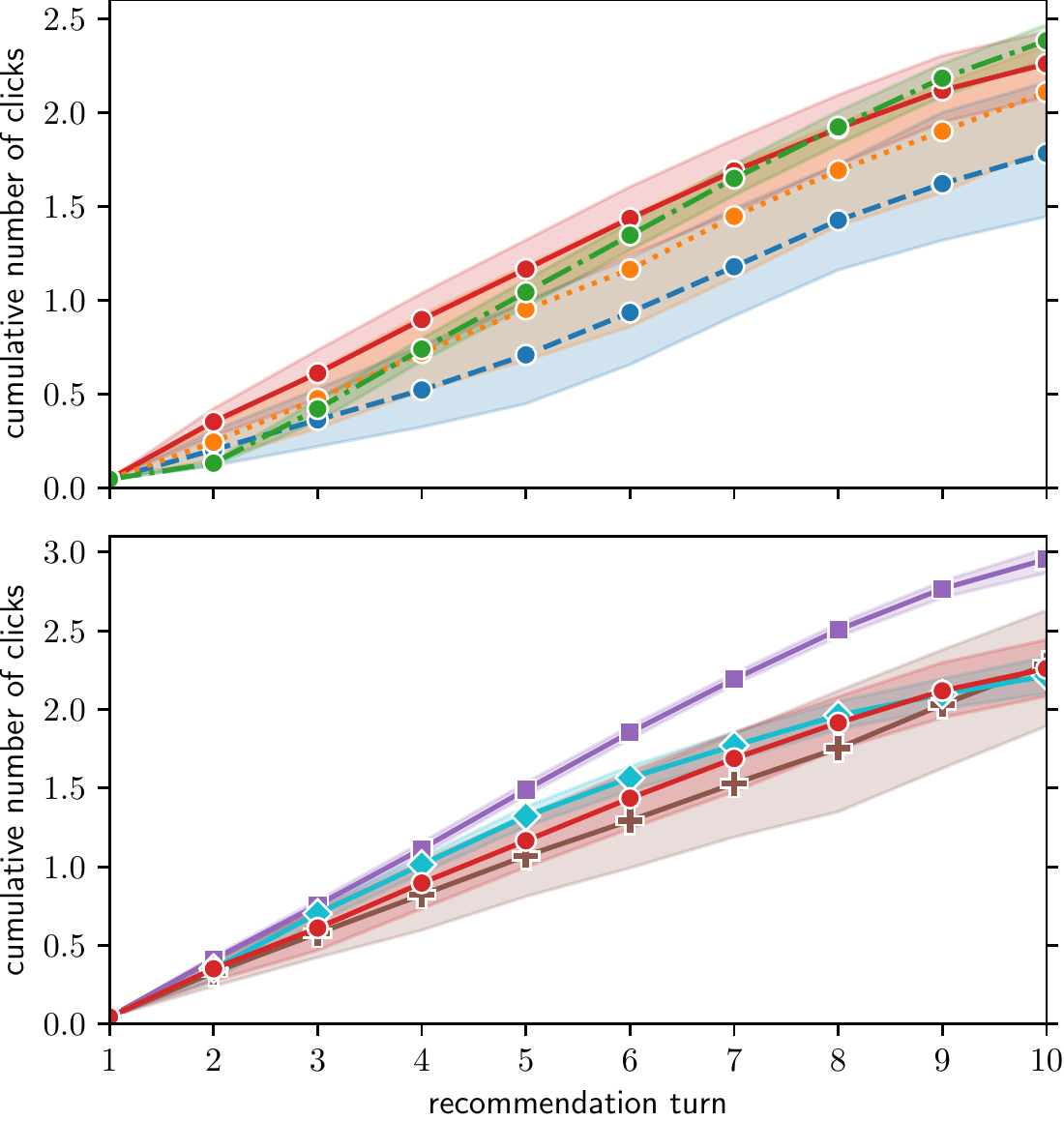}
	  \label{fig:eval_curves:coat}
	  } %
    \caption{Comparisons of evaluation performance (the cumulative number of clicks) among policies with four state encoders proposed by \citet{liu2020state} (top), and between attention and the additional \ac{MLP},\protect\footnotemark{} \ac{GRU}, and \ac{CNN}-based state encoders (bottom)
	on the unbiased simulations generated from the unbiased test sets of Yahoo!\ R3 and Coat shopping datasets, respectively.}
    \label{fig:eval_curves}
\end{figure*}

\section{Experimental Results and Analysis}

Our experiments results are meant to determine whether the main finding of \citet{liu2020state} can be reproduced:
\begin{quote}
\emph{The attention state encoder for \ac{RL4Rec} provides significantly higher performance than the \ac{BOI}, \ac{PLD} and \ac{Avg} state encoders.}
\end{quote}
Moreover, in our analysis we investigate whether this finding generalizes in the four directions described at the start of Section~\ref{sec:reproduce:comparison}.

\subsection{Comparison of state encoders on debiased simulation of Yahoo!\ R3 dataset}
We start our analysis by considering our first research question \ref{rq:reproduced}:
\emph{whether \citet{liu2020state}'s finding generalizes to DQN-based RL4Rec methods when evaluated in the debiased SOFA simulator and compared with more state encoders}.

Fig.~\ref{fig:eval_curves:yahoo} (top) displays the evaluation performance of the optimized policies based on four state encoders proposed by \citet{liu2020state}; the reported metric is the average cumulative number of clicks received over 10 interaction turns.
The first interaction turn is always represented by the empty state, and as a result, the choice of state encoder is inconsequential and the performance of all state encoders is identical.
As the number of interaction turns becomes larger, the differences between the state encoders become more apparent.
On the simulations of Yahoo!\ R3 dataset -- the same dataset used by~\citeauthor{liu2020state} --, as shown in Fig.~\ref{fig:eval_curves:yahoo} (top), we see results consistent with those reported by \citeauthor{liu2020state}:
\begin{enumerate*}
	\item \ac{BOI} and \ac{PLD} perform comparably and worse than \ac{Avg}; and
	\item on average the attention state encoder outperforms \ac{BOI}, \ac{PLD}, and \ac{Avg}.
\end{enumerate*}

\footnotetext{\label{fn:MLP:activation} For the \ac{MLP}-based state encoder, we use a ReLU for Yahoo!\ R3 and tanh for the Coat shopping dataset, which we found to be the optimal choices for the corresponding datasets.}

Fig.~\ref{fig:eval_curves:yahoo} (bottom) displays the evaluation performance of the attention, \ac{MLP},\footref{fn:MLP:activation} \ac{CNN} and \ac{GRU} state encoders on the Yahoo!\ R3 dataset. We see that on average the attention state encoder performs similarly to the \ac{GRU} state encoder over ten interaction turns and better than the \ac{CNN} state encoder.
Thus, we confirm that attention is the optimal choice in our experimental setting on the Yahoo!\ R3, the same dataset as used in the original comparison~\cite{liu2020state}.
The main difference between \ac{MLP} and \ac{GRU} is the recurrent nature of the latter, thus it is the likely reason for why \ac{GRU} outperforms \ac{MLP}.
Similarly, the higher performance of attention over \ac{GRU} must be because of the additional attention layer, as this is the sole difference between the two state encoders.

We answer \ref{rq:reproduced} in the affirmative: \citeauthor{liu2020state}'s finding regarding the superiority of using the attention state encoder generalizes to \ac{DQN}-based \ac{RL4Rec} methods when evaluating in the debiased \ac{SOFA} simulation based on the Yahoo!\ R3 dataset used by~\citeauthor{liu2020state}, and compared with three more state encoders,  \ac{MLP}, \ac{GRU}, and \ac{CNN}.

\subsection{Comparison on a different dataset}
Now that we have found \citeauthor{liu2020state}'s finding to be reproducible in a debiased simulation generated from the Yahoo!\ R3 dataset, we consider the second research question \ref{rq:dataset}:
\emph{whether it also generalizes to a debiased simulation based on a different dataset}.

Fig.~\ref{fig:eval_curves:coat} displays the performance of different state encoders on the debiased simulation based on the Coat shopping dataset, which was not part of the original comparison~\cite{liu2020state}.
We make two observations from the top part of Fig.~\ref{fig:eval_curves:coat}:
\begin{enumerate*}
	\item on average, \ac{PLD} performs better than \ac{BOI}, but worse than \ac{Avg};
	\item attention has worse performance than \ac{Avg} over 10 interaction turns.
\end{enumerate*}
Furthermore, in the bottom part of Fig.~\ref{fig:eval_curves:coat} we see that:
\begin{enumerate*}[resume]
       \item attention does not have better performance than the additional \ac{MLP}, \ac{CNN} and \ac{GRU} state encoders;
	\item on average, attention performs comparably with \ac{GRU} and \ac{CNN}, although \ac{CNN} does suffer from a much higher variance; 
	\item the \ac{MLP} state encoder outperforms other state encoders significantly.
\end{enumerate*}
Thus, in stark contrast with our results on the \text{Yahoo!\ R3} dataset, on the Coat shopping dataset we do not observe the attention state encoder to have the highest performance.

Two potential reasons for this observed inconsistency between the two datasets could be
\begin{enumerate*}
	\item the difference in size between the two datasets:
	in contrast to attention, the \ac{Avg} and \ac{MLP} methods with fewer parameters are possibly more effective on the smaller Coat shopping dataset; and
	\item the different recommendation scenarios: there could be a stronger dependency between items in user interactions in an online shopping scenario (Coat shopping) than in a music recommendation scenario (Yahoo!\ R3).
\end{enumerate*}

Therefore, we answer \ref{rq:dataset} negatively: \citeauthor{liu2020state}'s finding does \emph{not} generalize to the debiased simulation with a different dataset.
In particular, attention is \emph{not} the optimal choice of state encoder for \acp{RL4Rec} when evaluating in the Coat shopping dataset, which was not considered by~\citet{liu2020state}.

In addition, we also did not observe a consistent performance for the additional \ac{MLP}, \ac{CNN} and \ac{GRU} state encoders across the two datasets.
On the Yahoo!\ R3 dataset, \ac{GRU} performs best (out of the three) and \ac{MLP} performs worst; 
yet on the Coat shopping dataset \ac{GRU} performs similarly to \ac{CNN} but considerably worse than \ac{MLP}.
This observation suggests that the relative effectiveness of state encoders depends on the dataset to which they are applied. 
Importantly, there is no single optimal state encoder applicable to \acp{RL4Rec} for all datasets.

\subsection{Convergency of RL4Recs state encoders}
\begin{figure*}[t]
    \centering
	\subfigure{
	  \includegraphics[width=0.73\textwidth,trim=0 0 0 0]{image/plot_legend.pdf}
	  } \setcounter{subfigure}{0}
    \subfigure[Yahoo!\ R3.]{
		\includegraphics[width=.48\textwidth]{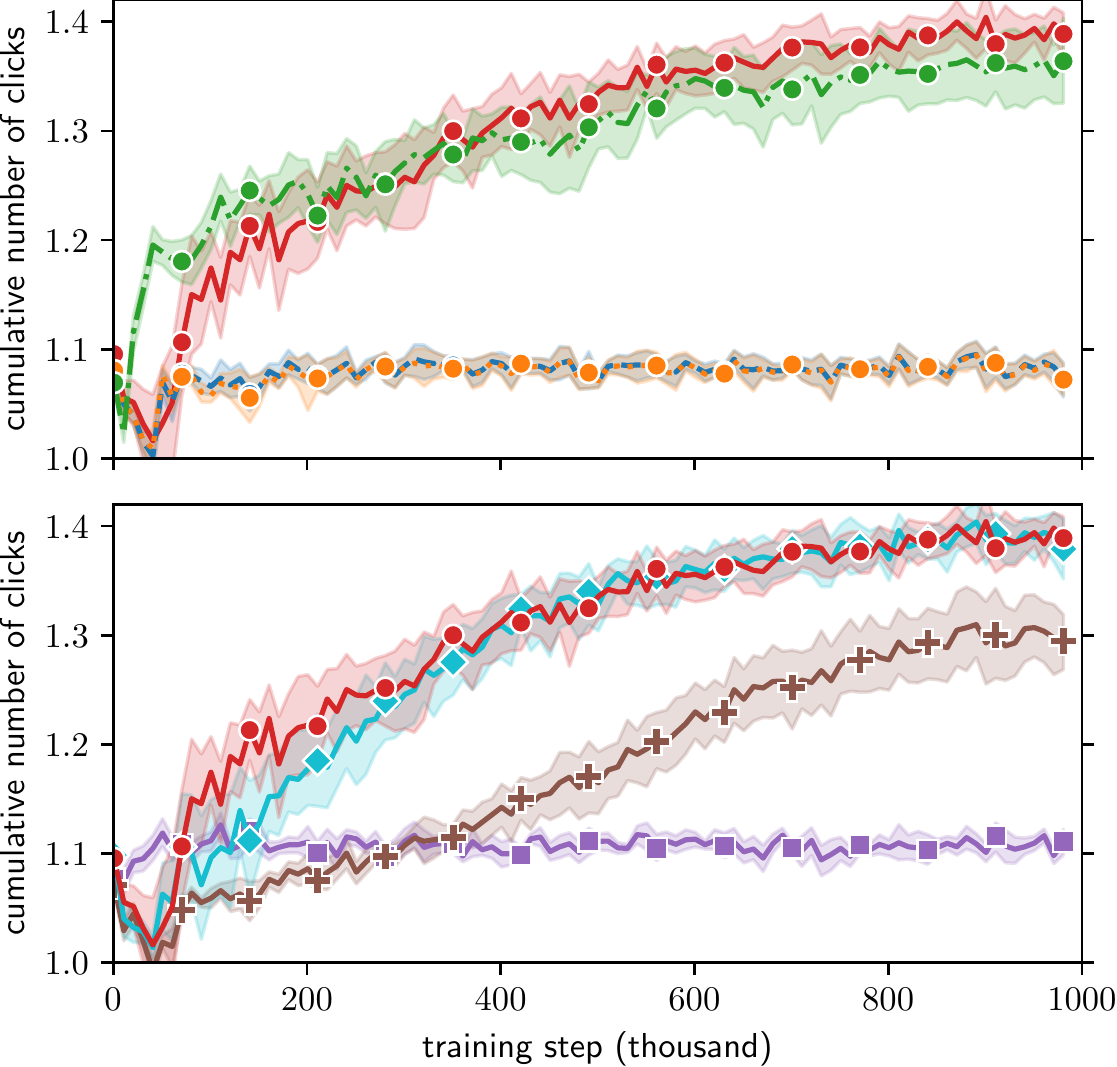}
		\label{fig:train_curves:yahoo}
	  }  
	\hspace{0.2cm}
	\subfigure[Coat shopping.]{
		\includegraphics[width=.48\textwidth]{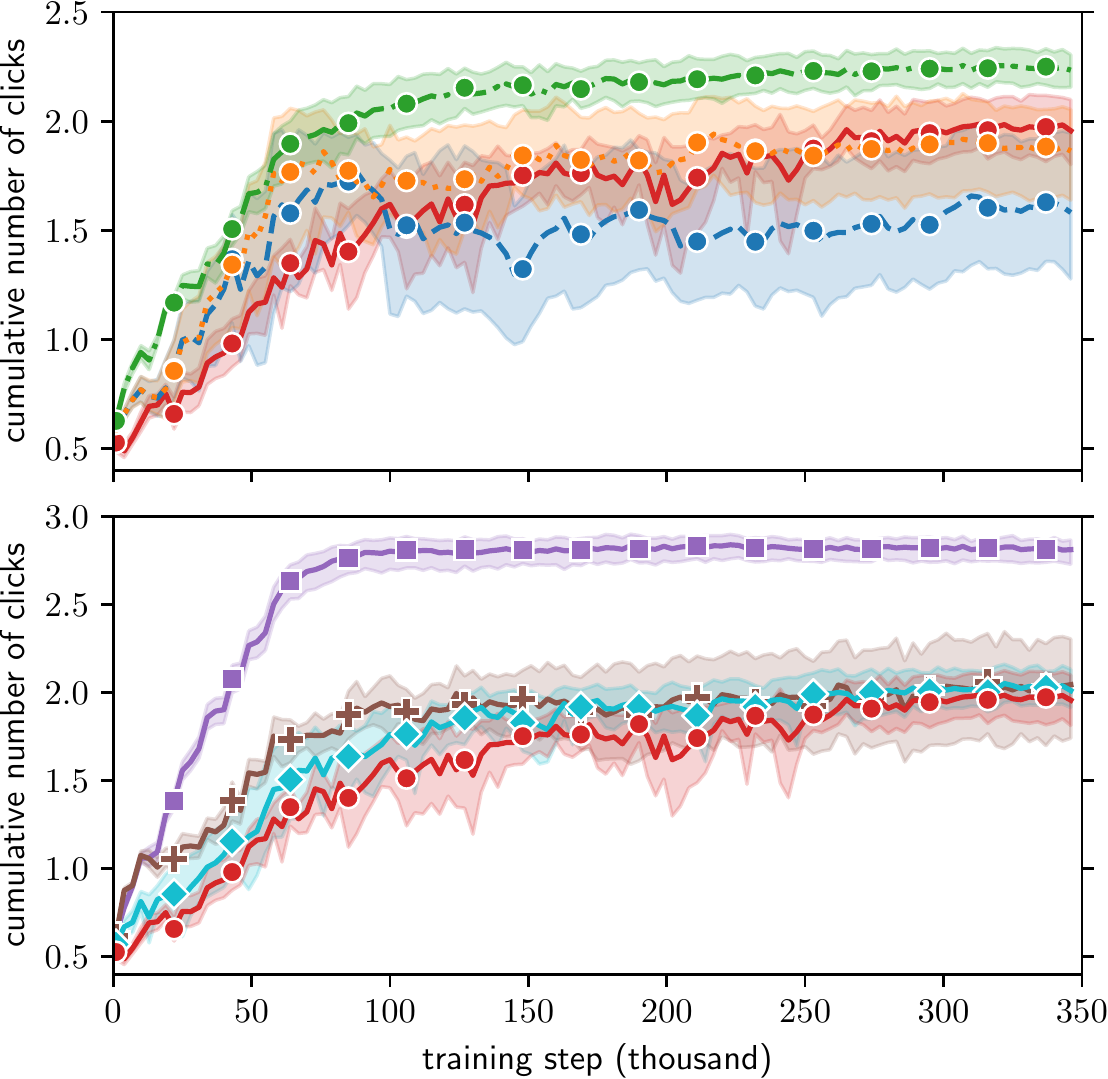}
		\label{fig:train_curves:coat}
	  } %
    \caption{Learning curves tracking average cumulative number of clicks received by policies with four state encoders (top), and with attention and additional \ac{MLP}, \ac{GRU} and \ac{CNN}-based state encoders (bottom) on the debiased simulations generated from training sets of Yahoo!R3 and Coat shopping datasets, respectively.}
    \label{fig:train_curves}
\end{figure*}

\begin{table}[t]
	\centering
        \caption{Training time in seconds for 1,000 training steps.} \vspace{1mm}
		\label{tab:training_time}%
			\begin{tabular}{l c c c c c c c}
			\toprule
            \bf Dataset & \bf BOI & \bf PLD & \bf Avg & \bf MLP & \bf CNN & \bf GRU & \bf Att\\
			\midrule
            Coat & 6.6 & 7.0 & 6.4 & 6.4 & 7.6 & 23.4 & 25.4 \\
            Yahoo!\ R3 & 9.0 & 9.6 & 8.4 & 9.4 & 10.8 & 26.8 & 30.4\\
			\bottomrule
		\end{tabular}%
\end{table}%
Convergence is also a crucial property for \ac{RL4Rec} methods because they are more prone to divergence problems as they continuously update recommendation policies while interacting with users.
Next, we investigate how the choice of state encoder affects the convergence of \ac{DQN}, in terms of the number of training steps and training time needed to converge.

Fig.~\ref{fig:train_curves} displays the learning curves of policies with different state encoders, which track the average cumulative number of clicks over 10 interaction turns on the debiased simulations on the \text{Yahoo!\ R3} and the Coat shopping datasets.
We observe that:
\begin{enumerate*}[label=(\arabic*)] 
	\item \ac{BOI} and \ac{PLD} converge the earliest but to policies that receive only a small cumulative number of clicks;
	\item \ac{MLP} has a similar convergence speed as \ac{BOI} and \ac{PLD} but its performance at convergence greatly varies between the datasets;
	\item \ac{MLP} converges faster than \ac{Avg}, suggesting that its activation function speeds up the learning process;
	\item attention converges slightly slower than \ac{GRU}, most likely due to having more parameters; and %
	\item the convergence speed of \ac{CNN} greatly varies between the two different datasets.
\end{enumerate*}
In summary, state encoders with few parameters, \eg \ac{Avg} and \ac{MLP}, converge faster than those with more parameters, \eg attention.

Furthermore,
Table~\ref{tab:training_time} clearly shows that the time for training on the larger Yahoo!\ R3 dataset is longer than on the Coat shopping dataset, which contains fewer items and users. 
As expected, \ac{Avg} and \ac{MLP} have the fewest parameters and accordingly also require less training time per thousand training steps.
\ac{BOI} and \ac{PLD} take slightly more time than \ac{Avg} which could be explained by the higher dimensionality of their state representations.
Lastly, attention is more time consuming than \ac{GRU}, which is likely due to its additional attention layer.
In summary, the attention state encoders require a higher computation cost, despite the fact that they do not always guarantee to reach the highest performance, \eg on the Coat shopping dataset. 

\subsection{Choice of activation functions for MLP}
\begin{figure*}[t]
    \centering
	\subfigure{
	  \includegraphics[width=0.65\textwidth,trim=0 0 0 0]{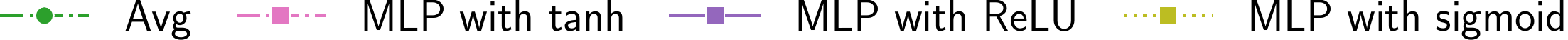}
	  } \setcounter{subfigure}{0}
    \subfigure[Yahoo!\ R3.]{
		\includegraphics[width=.48\textwidth]{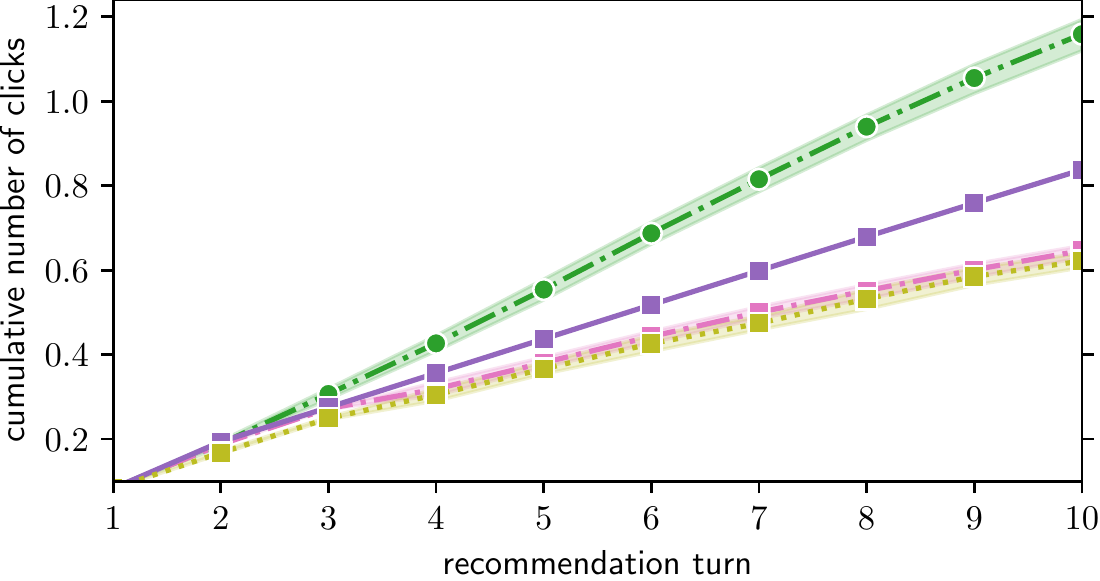}
		\label{fig:MLP:yahoo}
	  }  
	\hspace{0.2cm}
	\subfigure[Coat shopping.]{
		\includegraphics[width=.48\textwidth]{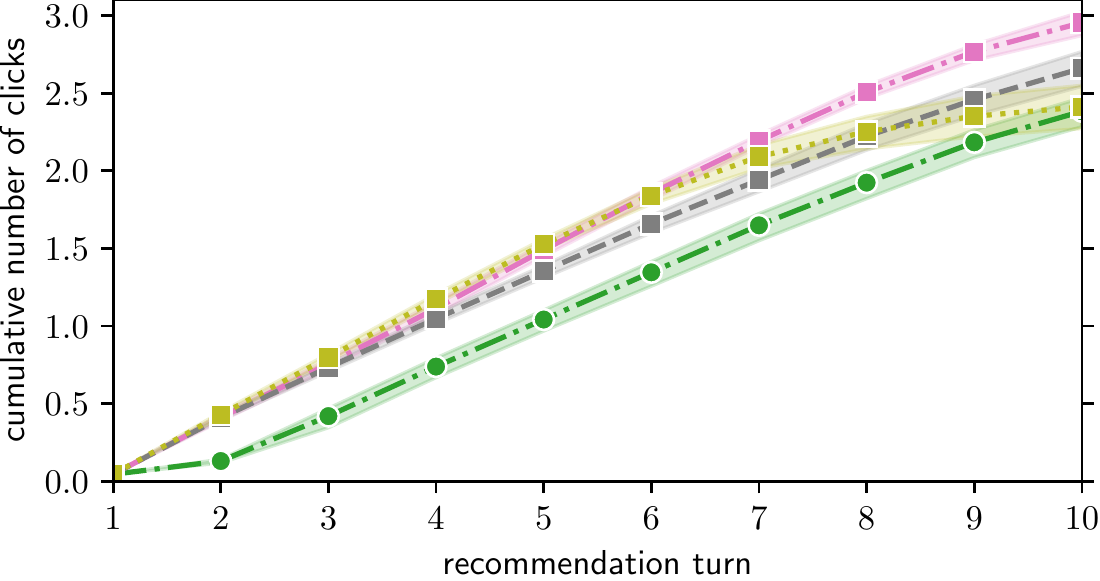}
		\label{fig:MLP:coat}
	  } %
    \caption{Evaluation performance (the cumulative number of clicks) of policies with Avg, and MLP state encoders with different activation functions (tanh, ReLU, and sigmoid)
	on the simulations of Yahoo!\ R3 and Coat shopping datasets, respectively.}
    \label{fig:MLP:activation}
\end{figure*}

The \ac{MLP} state encoders apply a non-linear activation function on top of \ac{Avg} and show varying evaluation performance when applied to different datasets: we have seen that it performs best on the Coat shopping dataset, but worse than \ac{Avg} on Yahoo!\ R3, as shown in Fig.~\ref{fig:eval_curves} (bottom row).
These observations prompt us to consider \ref{rq:activation}: 
\emph{whether the choice of activation functions should be taken into account when using the \ac{MLP}-based state encoder for \acp{RL4Rec}}.

Fig.~\ref{fig:MLP:activation} displays the comparison of evaluation performance between \ac{Avg} and \acp{MLP} with the tanh, ReLU, and sigmoid activation functions.
We observe that:
\begin{enumerate*}
	\item interestingly, \acp{MLP} perform better on the Yahoo!\ R3 dataset but worse than \ac{Avg} on the Coat shopping dataset; we speculate that this is due to the different sizes of the two datasets and the different recommendation scenarios they represent;
	\item for \acp{MLP}, sigmoid is the worst choice of activation function for simulations on both datasets, probably because it is more prone to the vanishing gradient problem~\cite{bengio1994learning}; and
	\item the performance with tanh and ReLU is not consistent across both datasets: tanh has the best performance on Coat, but is worse than ReLU on Yahoo!\ R3.
\end{enumerate*}

Therefore, we answer \ref{rq:activation} in the affirmative: the choice of activation function should certainly be taken into account when using \ac{MLP}-based state encoders for \ac{RL4Rec}.
Furthermore, our observations also suggest that the choice of activation function greatly depends on the dataset to which the \ac{MLP} state encoder will be applied.

\section{Conclusion}

In this paper, we have reproduced and generalized a previous study by \citet{liu2020state} regarding the choice of state encoder for \acfp{RL4Rec} in four directions:
\begin{enumerate*}
    \item a debiased simulated environment, named \ac{SOFA}; 
    \item \ac{RL4Rec} methods based on \acf{DQN}, the most popular \ac{RL} method used in \acp{RL4Rec};
    \item three additional state encoders based on three typical neural networks: \acfp{MLP}, \acfp{GRU}, and \acfp{CNN};
    \item besides the Yahoo!\ R3 dataset used in the original study~\cite{liu2020state}, we also considered the Coat shopping dataset as the basis for debiased simulations.
\end{enumerate*}
Our experimental results show that the higher performance of the attention state encoder over the \acf{BOI}, \acf{PLD}, and \acf{Avg} state encoders is reproducible in the debiased simulation generated from the Yahoo!\ R3 dataset, where \ac{DQN} was used instead of actor-critic \ac{RL}; moreover, the attention state encoder also outperforms the three additional \acf{MLP}, \acf{CNN} and \acf{GRU} state encoders on the debiased simulation based on the Yahoo!\ R3 dataset.
However, the attention state encoder performed worse than \ac{Avg} and \ac{MLP} when the simulation is based on the Coat shopping dataset, a dataset not used in~\citep{liu2020state}, despite the fact that it has the highest computational costs. 

In summary, our results confirm that \citeauthor{liu2020state}'s finding generalizes in the first three directions, \ie the debiased simulation, \ac{DQN}-based \ac{RL4Rec} method, and more state encoders, but does \emph{not} generalize to the debiased simulation generated from a different dataset, \ie the Coat shopping dataset.
In addition, we have found that the choice of activation function plays a crucial role when constructing a state encoder for \acp{RL4Rec}.

Future work should further investigate the importance of the choice of \ac{RL} methods for \ac{RL4Rec}.
A comparison of different \ac{RL} methods, such as \ac{DQN}, REINFORCE and actor-critic, in various \ac{RL4Rec} frameworks could reveal whether comparisons of \ac{RL} methods generalize across different settings.
The resulting insights could greatly aid researchers and practitioners in the \ac{RL4Rec} domain.

\section*{Implementation Resources and Data}
To facilitate the reproducibility of the reported results, this study only made use of publicly available data.
Our complete experimental implementation is publicly available with detailed instructions for reproducing our experiments at \url{https://github.com/BetsyHJ/RL4Rec}.

\section*{Acknowledgements}
This research was supported by the Hybrid Intelligence Center, a 10-year program funded by the Dutch Ministry of Education, Culture and Science through the Netherlands Organisation for Scientific Research, \url{https://hybrid-intelligence-centre.nl} and partially by the Google Research Scholar Program. 
All content represents the opinion of the authors, which is not necessarily shared or endorsed by their respective employers and/or sponsors.

\bibliographystyle{ACM-Reference-Format}
\balance
\bibliography{huang21}

\end{document}